\documentclass[a4paper,11pt]{article}
\pdfoutput=1 % if your are submitting a pdflatex (i.e. if you have
\usepackage{jinstpub} % for details on the use of the package, please
\usepackage{lineno}
\usepackage{float}
\title{\boldmath DIAMASIC: A multichannel front-end electronics for high-accuracy time measurements for diamond detectors}

\author[a,1]{A. Ghimouz,\note{Corresponding author.}}
\author[b]{F. Rarbi,}
\author[b]{O. Rossetto}
\affiliation[a]{Univ. Grenoble Alpes, Grenoble INP, CNRS, LPSC-IN2P3, 38000 Grenoble, France}
\affiliation[b]{Univ. Grenoble Alpes, CNRS, Grenoble INP, LPSC-IN2P3,38000 Grenoble, France}
\emailAdd{ghimouz@lpsc.in2p3.fr}

\abstract{This paper describes the design and testing results of an 8 channels preamplifier-discriminator circuit based on a resistive feedback Transimpedance Amplifier architecture and a Leading-Edge Discriminator stage for fast high-accuracy time measurement systems. The circuit has been designed in a $130 \; nm$ CMOS technology. It is intended to be used as a Front-End-Electronics for measuring the Time Of Flight using diamond detectors. The size of the chip is $1.27 \times 1.22 \; mm^2$ and the total power consumption of one channel is $1.5 \; mW$ with a power supply of $1.2 \; V$. Testing results shows a timing jitter of about $80 \; ps$ for a $10 \; fC$ input charge pulse.}


\collaboration[c]{on behalf of DIAMASIC project}

\proceeding{TWEPP 2021 Topical Workshop on Electronics for Particle Physics\\
  20–24 sept. 2021\\
  Online event}

\begin{document}
\maketitle
\flushbottom

\section{Introduction}
\label{sec:intro}

Recently, Chemical Vapor Deposition (CVD) Diamond detector offers an attractive alternative to silicon detector due to its outstanding performances such as the higher charge mobility ($1600 \; cm^2/Vs$ and $2100 \; cm^2/Vs$ for electron and hole respectively), low leakage current (high bandgap of $5.45 \; eV$) and radiation hardness capabilities~\cite{b01}. 

CVD diamond offers an attractive alternative to Si-PIN detector in nuclear physics experiment and pulse radiation research~\cite{b02}. This creates a need for dedicated Front-end Electronics (FEE) design that guarantees a minimum impact on the generated signals: optimum bandwidth, low noise and linearity.

This paper is organized as follows: Section II introduces the CVD diamond detectors. Section III describes details of the implementation of the FEE for high accuracy time measurements using the $g_m/I_d$ design methodology. Test results of the preamplifier's timing jitter performance  are presented in Section IV followed by the conclusion in Section V.

\section{Diamond detector}
\begin{figure}[H]
	\centering
	\def\svgwidth{1\columnwidth}
	\fontsize{10pt}{10pt}\selectfont\includegraphics[scale=1.6]{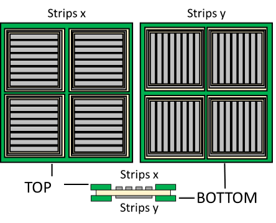}
	\caption{Concept diagram of the CVD double-side stripped metallized diamond detector.}
	\label{fig_1}
\end{figure}

The CVD diamond detector of this study is a double-side stripped metallized diamond used as a position sensitive detector. It is illustrated in figure~\ref{fig_1}.

In this case, diamonds are used as solid-state ionization chambers: the motion of the charges generated by the particles hitting the diamond creates an instantaneous current that induces on the electrodes connected to the strips. This means that the detector can be modeled as a current source with both a capacitor and a resistor in parallel of each other. The resistor is neglected because it is so high to be considered (several TΩ)~\cite{b03}.

%The segmentation of the detection surface into strips meets the need for two dimensional (2D) cross-sectional localization of the incident ion beam but also allows:
%\begin{itemize}
%\item A geometric definition of the lowest achievable spatial resolution, conditioned by the area defined by the intersection of two orthogonal strips;
%\item A distribution of the generated load on several reading channels, enables to improve the spatial resolution of the device to a value lower than the area described by the crossing of two orthogonal strips;
%\item A reduction in the size of the electrodes, thus minimizing the input capacitance of the read-out electronics.
%\end{itemize}

%\appendix
\section{Front-end electronics design}
We define a time measurement system as a FEE for radiation detectors that targets a time to digital converter (TDC)~\cite{b04}.% figure~\ref{fig_2} illustrates its different components.

%\begin{figure}[H]
	%\centering
	%\def\svgwidth{1\columnwidth}
	%\fontsize{10pt}{10pt}\selectfont\includegraphics[scale=0.5]{figure2.png}
	%\caption{Concept diagram of a solid state ionization chamber based on a diamond detector.}
	%\label{fig_2}
%\end{figure}

We focus on the input stage: the preamplifier, where we study the effect of its features on the final timing resolution. This latter is mathematically modeled as the maximum error of time defined using equation~\ref{115}:

\begin{equation}
\sigma_t = \sqrt{\sigma_{Ji}^2+\sigma_{TW}^2+\sigma_{TDC}^2}
\label{115}
\end{equation}
Where:
\begin{itemize}
\renewcommand\labelitemi{--}
\item $\sigma_{TDC}$ is linked to the resolution of the TDC which depends on its architecture;
\item $\sigma_{TW}$ is related to the discrimination stage and known as the time walk;
\item $\sigma_{Ji}$ is usually associated to the noise of the preamplifier stage.
\end{itemize}

The $\sigma_{Ji}$ represents the most critical element which we need to reduce  to achieve high timing resolutions ($<100 \; ps$). During this study, this error will be considered as the criteria of the time resolution.

\subsection{The $g_m/I_d$ design methodology}

\begin{figure}[H]
	\centering
	\def\svgwidth{1\columnwidth}
	\fontsize{10pt}{10pt}\selectfont\includegraphics[scale=0.7]{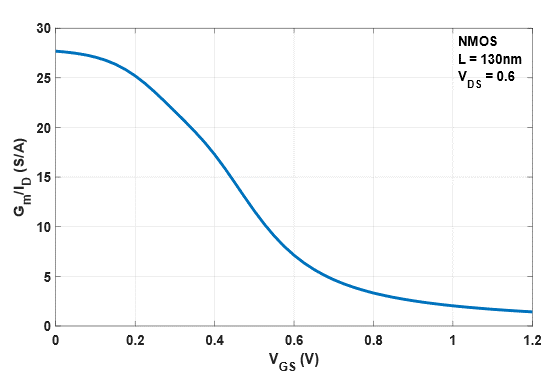}
	\caption{The extracted $g_m/I_d$ values for different values of $V_{GS}$ of our used 130nm CMOS technology.}
	
	\label{fig_3}
\end{figure}

The $g_m/I_d$ design methodology allows to captures the relation between the fundamental function of the transistor (because fundamentally, a transistor is a voltage controlled current source), which is its transconductance $g_m$ and its power consumption $I_d$. It was shown in~\cite{b05} that the $g_m/I_d$ ratio is directly related to the most important analog specifications: speed, noise, efficiency, gain, swing and mismatch.  Moreover, the range of values of $g_m/I_d$ is very limited, typically between 0 to 30. We illustrate its values after we extracted the LUTs of the used technology (130 nm CMOS technology)in figure~\ref{fig_3}. Also, it is important to know that this does not differs much from a device to another and from a technology to another~\cite{b10}.
 
\begin{figure}[H]
	\centering
	\def\svgwidth{1\columnwidth}
	\fontsize{10pt}{10pt}\selectfont\includegraphics[scale=0.6]{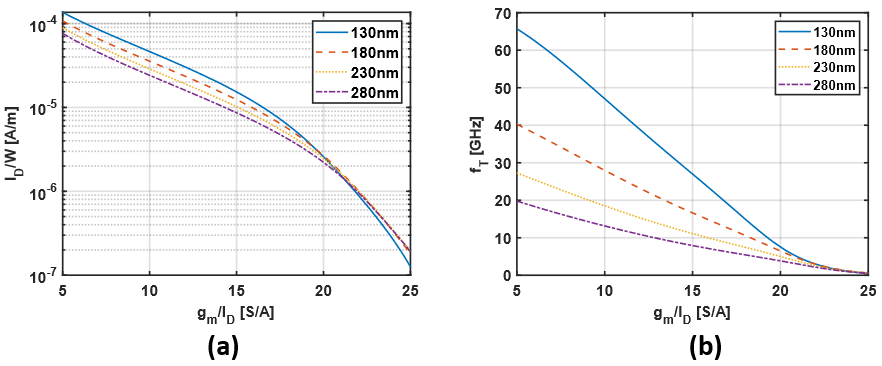}
	\caption{Performance of (a) power consumption, (b) speed against the The extracted $g_m/I_d$ values of our used 130nm CMOS technology.}
	
	\label{fig_4}
\end{figure}

%\FloatBarrier
The use of the $g_m/I_d$ design methodology guarantee an efficient sizing of the transistors of the chosen topology, we apply it through a model-based design approach using MATLAB. In figure~\ref{fig_4} (a) we can see that for the chosen technology (130nm CMOS), we can achieve a low power design while placing our transistors in a weak inversion mode. This ensures the lowest possible consumption.  It illustrates as well (figure~\ref{fig_4} (b)) that we keep a sufficient speed for such choice.
 
\subsection{TIA topology}
\begin{figure}[H]
	\centering
	\def\svgwidth{1\columnwidth}
	\fontsize{10pt}{10pt}\selectfont\includegraphics[scale=0.35]{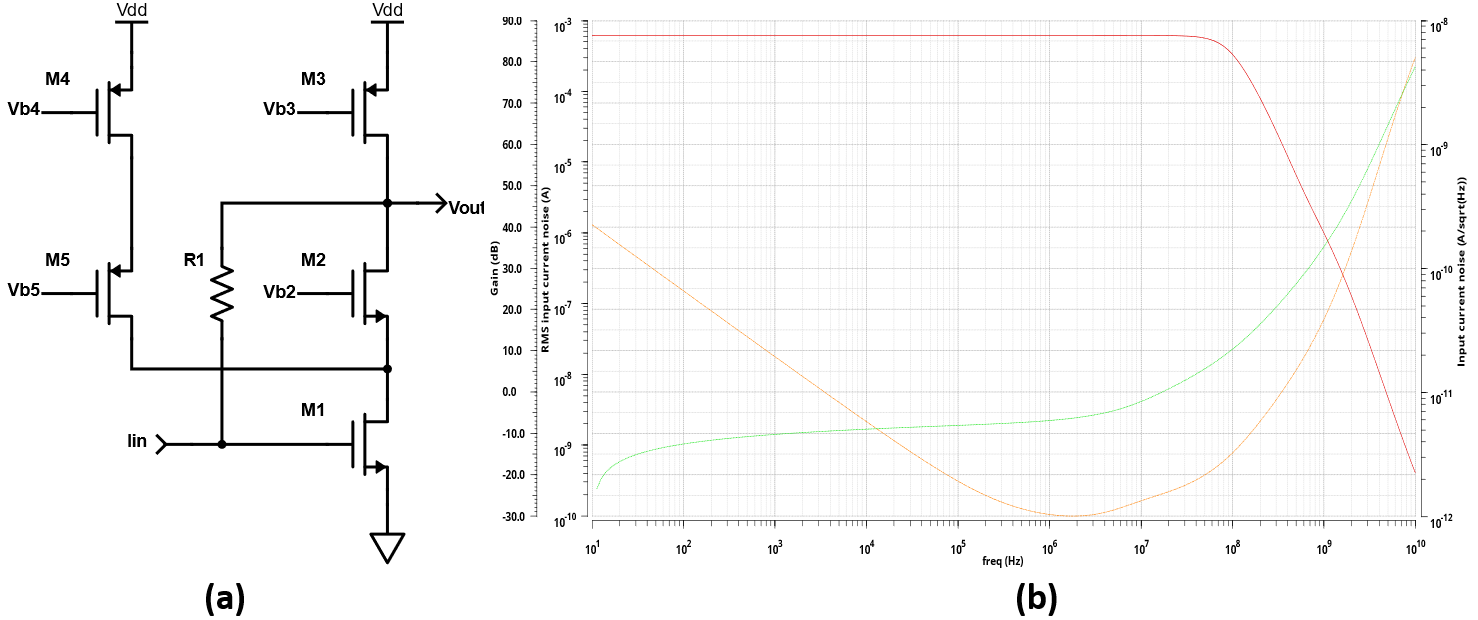}
	\caption{The extracted $g_m/I_d$ values for different values of $V_{GS}$ of our used 130nm CMOS technology.}
	
	\label{fig_5}
\end{figure}

The proposed circuit is an 8 channels preamplifier-discriminator circuit based on a resistive feedback Transimpedance Amplifier (TIA) architecture shown in figure~\ref{fig_5} (a) and a Leading-Edge Discriminator stage. For this TIA the gain is obtained directly from the feedback resistor. The use of the boost branch allows to minimize the current of the load of the main node and thus a higher open loop gain. The cascode helps to decrease the miller effect in order to preserve the targeted bandwidth.

With this design, we achieve a DC gain of 87 dB and an UGF of 5 GHZ with an equivalent RMS current input noise of around 50nA as illustrated in figure~\ref{fig_5} (b). 
%\begin{figure}[H]
	%\centering
	%\def\svgwidth{1\columnwidth}
	%\fontsize{10pt}{10pt}\selectfont\includegraphics[scale=0.55]{figure6.png}
	%\caption{The extracted $g_m/I_d$ values for different values of $V_{GS}$ of our used 130nm CMOS technology.}
	
\begin{figure}[H]
	\centering
	\def\svgwidth{1\columnwidth}
	\fontsize{10pt}{10pt}\selectfont\includegraphics[scale=0.33]{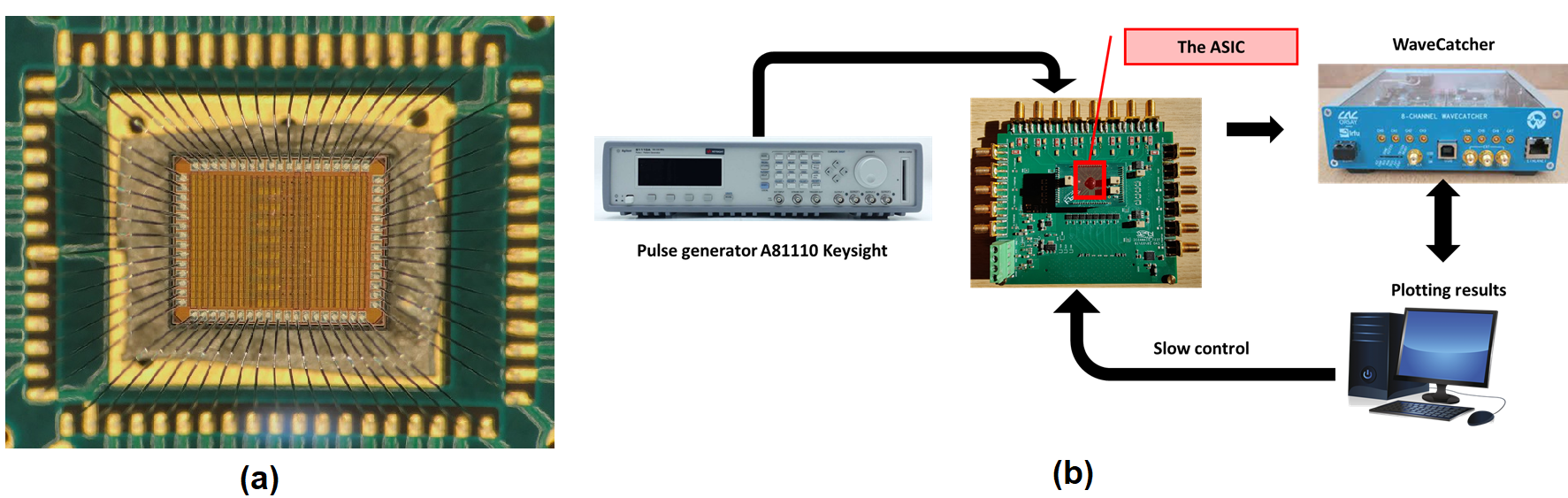}
	\caption{(a)The ASIC of the proposed FEE, (b) Testing setup}
	
	\label{fig_7}
\end{figure}

	%\label{fig_6}
%\end{figure}
The circuit has been fabricated in a $130 \; nm$ CMOS technology. The size of the chip is $1.27 \times.22 \;mm^2$ and the total power consumption of one channel is $1.5 \; mW$ with a power supply of $1.2 \; V$. Figure~\ref{fig_7} (a) shows the fabricated circuit.

\section{Test results}
For the tests, the chip is mounted directly on a PCB (both input and output are bonded). This PCB is mounted on a test card where the supply and bias are applied. We emulate the behavior of the detector by injecting different values of charges through a capacitor of 1 pF using a precise pulse generator as shown in figure~\ref{fig_7} (b). All the channels are activated. We record the values of the output for 1000 tests with every input in order to estimate as precisely as possible the timing jitter. Here we show the results in the case of an input charge of 10 fC, we achieve a timing jitter of around 85 ps. This value decreases for higher input charges as shown below.

\begin{figure}[H]
	\centering
	\def\svgwidth{1\columnwidth}
	\fontsize{10pt}{10pt}\selectfont\includegraphics[scale=0.52]{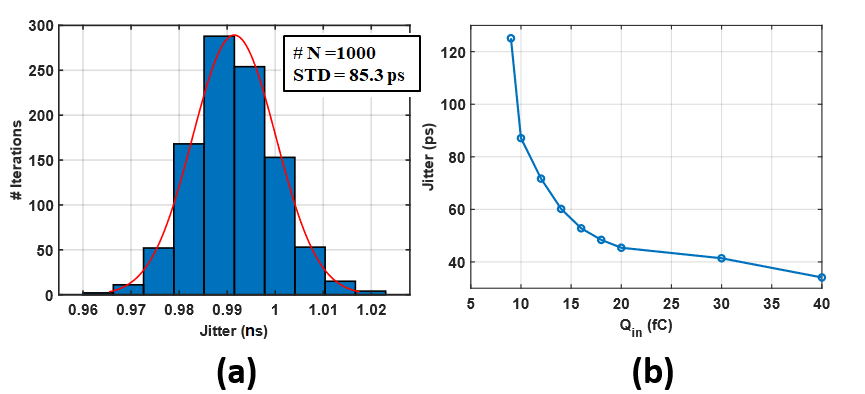}
	\caption{Measurement results: (a) Measured timing jitter for a 10 fC input charge 
, (b) Measured timing jitter for  different input charges.}
	
	\label{fig_8}
\end{figure}

\section{Conclusion}
In this paper, we discussed the design and testing results of an 8 channels preamplifier-discriminator circuit based on a resistive feedback Transimpedance Amplifier architecture and a Leading-Edge Discriminator stage for fast high-accuracy time measurement systems. The size of the chip is $1.27 \times 1.22 \; mm^2$ and the total power consumption of one channel is $1.5 \; mW$ with a power supply of $1.2 \; V$. Testing results shows a timing jitter of about $80 \; ps$ for a $10 \; fC$ input charge pulse.

%\acknowledgments

%This is the most common positions for acknowledgments. A macro is
%available to maintain the same layout and spelling of the heading.

%\paragraph{Note added.} This is also a good position for notes added
%after the paper has been written.

% We suggest to always provide author, title and journal data:
% in short all the informations that clearly identify a document.

\end{document}